\documentclass[aps,prl,onecolumn,superscriptaddress,groupedaddress]{revtex4}
\usepackage{subfig}
\usepackage{graphicx}  
\usepackage{dcolumn}   
\usepackage{bm}        
\usepackage{amssymb}   
\usepackage{slashed}
\usepackage{graphicx}				
\usepackage{amsmath}
\usepackage{mathtools}
\usepackage{tikz,pgf}
\usepackage{comment}
\usepackage{asymptote}
\usetikzlibrary{arrows,backgrounds}
\usetikzlibrary{fit,scopes,calc,matrix,positioning,decorations.pathmorphing}
\usepackage[all]{xy}
\usepackage{yfonts}
\newcommand{\bra}[1]{\ensuremath{\left\langle#1\right|}}
\newcommand{\ket}[1]{\ensuremath{\left|#1\right\rangle}}

\begin{document}
\title{Quantum Coherence and Chaotic Dynamics: Guiding Molecular Machines Toward Low-Entropy States}
\author{Andrei T. Patrascu}
\address{FAST Foundation, Destin FL, 32541, USA\\
email: andrei.patrascu.11@alumni.ucl.ac.uk}
\begin{abstract}
Quantum coherence profoundly alters classical thermodynamic expectations by modifying the structure and accessibility of probability distributions. Classically, transitions to lower-entropy states (local second-law violations) are exponentially suppressed, as lower-entropy configurations have fewer available microstates and are statistically improbable. However, introducing quantum coherence and structured quantum interference among semiclassical trajectories significantly changes this scenario. Quantum coherence reduces local entropy by establishing correlations among states that are classically independent, effectively restructuring probability amplitudes to channel transitions toward otherwise improbable low-entropy states. We analyze this phenomenon explicitly within the framework of semiclassical approximations, employing the Van Vleck-Gutzwiller propagator to quantify how interference terms arising from coherent superpositions modify classical fluctuation theorems. We demonstrate that quantum coherence, especially when combined with chaos-assisted dynamical tunneling, greatly enhances the probability of transitions to low-entropy configurations, creating pronounced counter-ergodic effects. Furthermore, by considering purification of mixed quantum states, we propose methodologies for deliberately engineering quantum phases among interfering pathways. This phase engineering can explicitly enhance coherence-driven transitions into lower-entropy states, providing a novel thermodynamic resource. Finally, we explore the feasibility of molecular-scale quantum machines exploiting these principles, highlighting their potential applications in quantum thermodynamics and quantum biology for performing useful work through coherence-mediated entropy reduction.
\end{abstract}
\maketitle
\section{Introduction}
Quantum coherence significantly reshapes classical thermodynamic expectations by altering the fundamental structure and accessibility of probability distributions. Classically, entropy-decreasing transitions - those representing local violations of the second law - are highly improbable due to the limited number of accessible microstates associated with lower-entropy configurations [1-4]. Quantum mechanically, however, coherence and structured interference between semiclassical paths [5-8] create new pathways to these improbable states, fundamentally altering thermodynamic behaviour [9-11].

The semiclassical analysis of quantum thermodynamics typically employs the Van Vleck-Gutzwiller (VVG) propagator [12-14], a powerful tool bridging classical and quantum dynamics. The VVG propagator approximates quantum amplitudes by summing over classical trajectories, each weighted by its classical action and a prefactor determined by second derivatives of the action. The action differences among interfering classical paths yield quantum interference patterns that dramatically impact transition probabilities [15-17]. Traditionally, the VVG propagator also includes the Maslov index [18], accounting for phase shifts associated with conjugate points along trajectories. However, alternative semiclassical formulations exist, such as the Herman-Kluk propagator [19], which avoid the explicit calculation of a Maslov index by representing quantum states through coherent-state basis transformations and Gaussian wavepackets. Such approaches offer computational and conceptual advantages, particularly in systems exhibiting chaotic dynamics.

Coherence-induced correlations among quantum states play a crucial role in entropy reduction. Classically, entropy measures the number of accessible independent microstates, reflecting uncertainty or ignorance about the system's precise configuration. Quantum entropy, however, also incorporates coherence and structured interference effects, capturing not only the count of microstates but also their accessibility and the structured probabilities arising from quantum interference [20]. Coherence reduces entropy by correlating otherwise independent states, effectively decreasing the number of independent states required to describe the system fully. This structured coherence creates targeted probability flows, enabling quantum systems to reliably access low-entropy states that would remain highly improbable in classical statistical frameworks.

In this article, we examine how quantum coherence and interference, particularly when combined with chaos-assisted dynamical tunnelling [21-23], enhance the probability of transitions to lower-entropy states. We show that coherence-induced correlations among trajectories reduce local entropy, channeling quantum amplitudes toward otherwise improbable low-entropy configurations. Furthermore, we explore the role of quantum purification - a method of embedding mixed quantum states into larger Hilbert spaces - to precisely control phase relationships among quantum pathways. Such phase engineering techniques can actively guide transitions into states of lower entropy, providing novel mechanisms for thermodynamic control.

We conclude by considering the practical implications and feasibility of molecular-scale quantum machines that exploit these quantum coherence principles. Such devices hold promise not only in advancing theoretical understanding but also in inspiring novel applications in quantum thermodynamics and quantum biology, where structured coherence could enable new mechanisms for efficient energy conversion and information processing.

\section{Quantum Modification of the Fluctuation Theorem via Van Vleck-Gutzwiller Propagator}

In classical thermodynamics, the fluctuation theorem quantifies the ratio of probabilities of observing entropy-increasing versus entropy-decreasing transitions as follows:
\begin{equation}
\frac{P(+\Delta S)}{P(-\Delta S)} = e^{\Delta S/k_B}.
\end{equation}
This expression emerges directly from classical trajectories, each considered independently. Quantum mechanically, however, transition probabilities must incorporate quantum coherence and interference effects, fundamentally modifying this classical relationship.

Semiclassically, quantum transition amplitudes are approximated using the Van Vleck-Gutzwiller propagator, expressed as:
\begin{equation}
K(q_f,q_i;t) \approx \sum_{\alpha}\sqrt{\frac{1}{2\pi i \hbar}\left|\frac{\partial^2 S_{\alpha}(q_f,q_i;t)}{\partial q_i \partial q_f}\right|} e^{\frac{i}{\hbar}S_{\alpha}(q_f,q_i;t)-\frac{i\pi}{2}\nu_{\alpha}},
\end{equation}
where $S_{\alpha}(q_f,q_i;t)$ is the classical action along trajectory $\alpha$ connecting initial position $q_i$ to final position $q_f$, and $\nu_{\alpha}$ is the Maslov index.

The probability of transitioning from state $A$ to state $B$ is thus given by the squared magnitude of the propagator:
\begin{equation}
P_{\text{qm}}(A\rightarrow B) = \left|\sum_{\alpha} A_{\alpha} e^{iS_{\alpha}/\hbar}\right|^2,
\end{equation}
where the amplitude $A_{\alpha}$ includes the prefactors from the Van Vleck-Gutzwiller expression. Crucially, this quantum mechanical probability includes interference terms, which do not appear classically:
\begin{equation}
P_{\text{qm}}(A\rightarrow B)=\sum_{\alpha}|A_{\alpha}|^2 + \sum_{\alpha\neq\beta} A_{\alpha} A_{\beta}^* e^{i(S_{\alpha}-S_{\beta})/\hbar}.
\end{equation}

When explicitly considering forward (entropy-increasing) and backward (entropy-decreasing) processes, the quantum modified fluctuation theorem becomes:
\begin{equation}
\frac{P_{\text{qm}}(+\Delta S)}{P_{\text{qm}}(-\Delta S)}=e^{\Delta S/k_B}\frac{1+\sum_{\alpha\neq\beta}\frac{A_{\alpha}A_{\beta}^*e^{i(S_{\alpha}-S_{\beta})/\hbar}}{\sum_{\alpha}|A_{\alpha}|^2}}{1+\sum_{\gamma\neq\delta}\frac{B_{\gamma}B_{\delta}^*e^{i(S_{\gamma}-S_{\delta})/\hbar}}{\sum_{\gamma}|B_{\gamma}|^2}}.
\end{equation}

Intuitively, quantum interference terms arising from action differences  create oscillatory corrections to classical probabilities. These corrections can strongly enhance the probability of entropy-decreasing transitions, especially when coherence-induced interference is constructive for such processes. Chaos-assisted dynamical tunnelling further amplifies these quantum corrections by connecting classically isolated states through coherent tunnelling paths mediated by chaotic trajectories.

Thus, the semiclassical quantum modification of the fluctuation theorem demonstrates that quantum coherence significantly increases the accessibility of entropy-decreasing transitions, fundamentally altering classical thermodynamic expectations. This result has profound implications for designing quantum devices and molecular systems that harness coherence-driven thermodynamic processes.

\section{Quantum Purification, Phase Engineering, and Enhanced Entropy-Decreasing Transitions}
Quantum purification is the process of embedding a mixed quantum state into a larger Hilbert space to form a pure state. Given a mixed state of a quantum system $\rho_{S}$:
\begin{equation}
\rho_{S}=\sum_{i}p_{i}\ket{s_{i}}\bra{s_{i}}
\end{equation}
with probabilities $p_{i}$ and orthonormal basis states $\ket{s_{i}}$, purification introduces an ancillary system $A$ to form a pure state:

\begin{equation}
\ket{\Psi_{SA}}=\sum_{i}\sqrt{p}\ket{s_{i}}\ket{a_{i}}
\end{equation}
where the ancilla states $\ket{a_{i}}$ constitute an orthonormal basis in the ancillary Hilbert space $\mathcal{H}$.

To achieve phase engineering, we introduce tunable relative phases $\phi_{i}$ into the purified state:
\begin{equation}
\ket{\Psi_{SA}(\phi)}=\sum_{i}\sqrt{p_{i}}e^{i\phi_{i}}\ket{s_{i}}\ket{a_{i}}
\end{equation}
The density matrix describing the purified state is thus:
\begin{equation}
\rho_{SA}\ket{\Psi_{SA}(\phi)}\bra{\Psi_{SA}(\phi)}=\sum_{i,j}\sqrt{p_{i}p_{j}}e^{i(\phi_{i}-\phi_{j})}\ket{s_{i}}\ket{a_{i}}\bra{s_{j}}\bra{a_{j}}
\end{equation}
The transition probability between states under unitary evolution $U$ is:
\begin{equation}
P(\Psi\rightarrow \Phi)=\lvert\bra{\Phi_{SA}}U\ket{\Psi_{SA}(\phi)}\rvert^{2}=\Bigg|\sum_{i}\sqrt{p_{i}}e^{i\phi_{i}}\bra{\Phi_{SA}}U\ket{s_{i}}\ket{a_{i}}\Bigg|^{2}
\end{equation}
Expanding this explicitly yields:
\begin{equation}
P(\Psi\rightarrow \Phi)=\sum_{i,j}\sqrt{p_{i}p_{j}}\bra{\Phi_{SA}}U\ket{s_{i}}\ket{a_{i}}\bra{s_{j}}\bra{a_{j}}U^{\dagger}\ket{\Phi_{SA}}
\end{equation}
Through precise tuning of the phases $\phi_{i}$, one can maximise constructive interference toward specific low-entropy target states, thereby significantly enhancing their transition probabilities. This targeted approach leverages quantum coherence and entanglement to systematically access and stabilise lower-entropy configurations, offering unprecedented control over thermodynamic processes at the quantum scale.
\begin{figure}[h!]
  \includegraphics[scale=0.5]{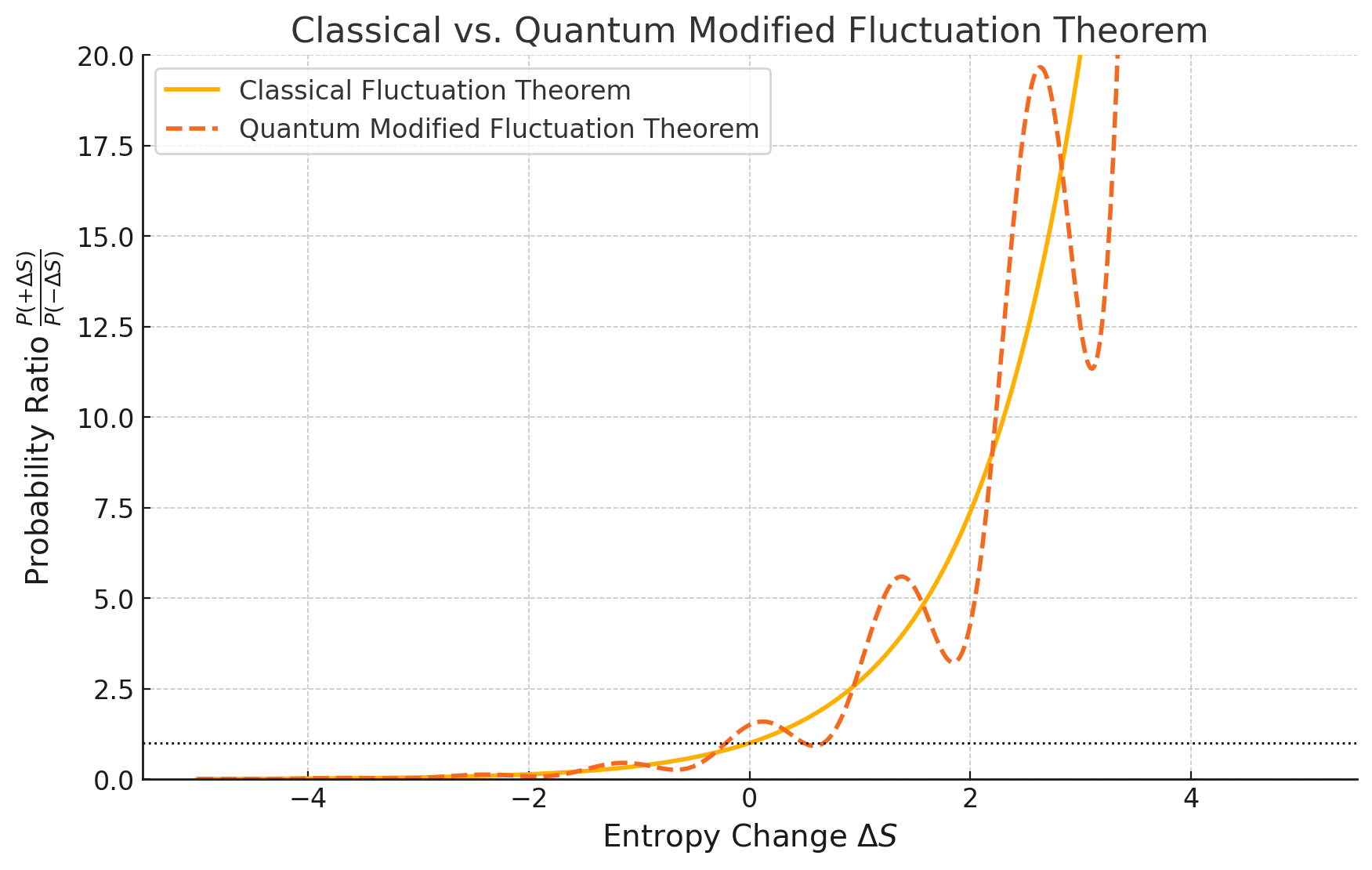}
  \caption{}
  \label{fig 1}
\end{figure}
The graphical representation illustrates the classical fluctuation theorem (solid line), which is an exponential function of entropy change, compared to the quantum-modified fluctuation theorem (dashed line). The quantum modification introduces oscillations due to quantum interference effects, significantly altering the probability ratio for entropy-increasing versus entropy-decreasing transitions.

In detail, the classical fluctuation theorem is expressed mathematically as:
\begin{equation}
\frac{P(+\Delta S)}{P(-\Delta S)}=e^{\frac{\Delta S}{k_{B}}}
\end{equation}
Quantum coherence modifies this relationship by introducing interference terms. The quantum-modified ratio can be written generally as:
\begin{equation}
\frac{P_{qm}(+\Delta S)}{P_{qm}(-\Delta S)}=e^{\frac{\Delta S}{k_{B}}}\frac{1+\sum_{\alpha\neq\beta}\frac{A_{\alpha}A_{\beta}^{*}e^{i(S_{\alpha}-S_{\beta})/\hbar}}{\sum_{\alpha}|A_{\alpha}|^{2}}}{1+\sum_{\gamma\neq\delta}\frac{B_{\gamma}B_{\delta}^{*}e^{i(S_{\gamma}-S_{\delta})/\hbar}}{\sum_{\gamma}|B_{\gamma}|^{2}}}
\end{equation}
Here $S_{\alpha},\;S_{\beta},\;S_{\gamma},\;S_{\delta}$ represent actions of interfering semiclassical trajectories, and the prefactors $A_{\alpha},\;A_{\beta},\;B_{\gamma},\;B_{\delta}$ come from the Van-Vleck-Gutzwiller propagator. The oscillatory behaviour introduced by quantum coherence clearly highlights enhanced probabilities of entropy-decreasing transitions, especially where constructive quantum interference occurs. This demonstrates how quantum interference significantly affects classical thermodynamic predictions. 

In the second graphical representation, the structured quantum coherence (orange dashed line) demonstrates a significant enhancement of transition probabilities specifically toward negative entropy changes, thus guiding the system into lower entropy states more effectively than the classical scenario (solid line).

Mathematically, this structured coherence is modelled as an interference pattern designed explicitly to amplify transitions toward targeted low-entropy states. A general expression capturing this concept could be represented as:
\begin{equation}
\frac{P_{structured}(+\Delta S)}{P_{structured}(-\Delta S)}=e^{\frac{\Delta S}{k_{B}}}[1+C e^{-\frac{(\Delta S-\Delta S_{0})^{2}}{2\sigma^{2}}}]
\end{equation}

\begin{figure}[h!]
  \includegraphics[scale=0.5]{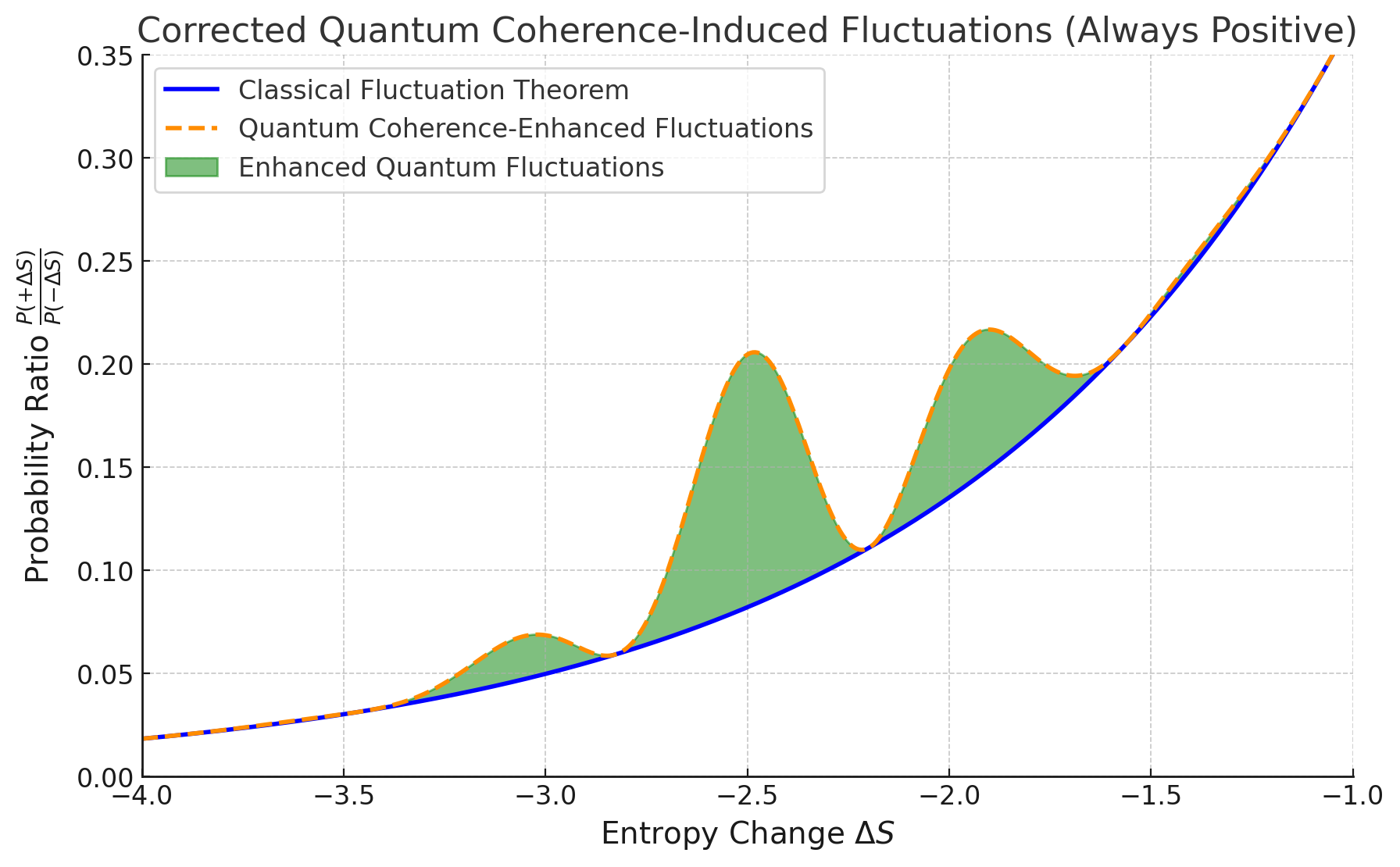}
  \caption{}
  \label{fig. 2}
\end{figure}

Here:
\begin{itemize}
\item $C$ is the strength of the coherence-induced enhancement.
\item $\Delta S_{0}$ specifies the targeted entropy reduction.
\item $\sigma$ determines the width of this coherence-driven enhancement.
\end{itemize}
In the shown example, the coherence effect was specifically chosen as a Gaussian centred at negative entropy change, significantly increasing the probability of transitions to low-entropy states (local second-law violations). Such a structured quantum coherence mechanism illustrates the fundamental ability of quantum systems to selectively channel probabilities, effectively overriding classical statistical constraints and thermodynamically favouring typically improbable transitions. 

\section{Role of Chaotic Dynamics in Semiclassical Entropy Reduction}

Chaos, characterised by extreme sensitivity to initial conditions, plays a crucial role in quantum systems within the semiclassical limit. Classically chaotic dynamics is associated with trajectories that diverge exponentially with time, filling the accessible phase space uniformly due to ergodicity. However, quantum mechanically, chaotic dynamics manifests through intricate interference patterns and coherence effects. In the semiclassical regime, these quantum features significantly modify the classical chaotic landscape.

Semiclassically, chaotic systems are described using multiple classical trajectories contributing to the propagator. The Van Vleck–Gutzwiller propagator, in particular, explicitly sums over a large set of trajectories:

\begin{equation}
K(q_{f},q_{i};t)\sim \sum_{\alpha}A_{\alpha}e^{iS_{\alpha}/\hbar}
\end{equation}

where each trajectory $\alpha$ contributes a phase determined by its classical action $S_{\alpha}$. In chaotic systems, slight variations in initial conditions produce significantly different trajectories, resulting in highly complex interference patterns that profoundly impact transition probabilities.

Quantum chaos introduces "chaos-assisted tunnelling," a phenomenon where quantum states associated with classically isolated regions in phase space become effectively coupled through chaotic trajectories. This tunnelling significantly enhances transitions to otherwise inaccessible or improbable states, particularly those with lower entropy.

Mathematically, chaos-assisted tunnelling probabilities are governed by interference terms of the form:
\begin{equation}
P_{tunnelling}\sim \Bigg|\sum_{\alpha,\beta}A_{\alpha}A_{\beta}^{*}e^{i(S_{\alpha}-S_{\beta}/\hbar}\Bigg|^{2}
\end{equation}
where pairs of trajectories $(\alpha,\beta)$ represent chaotic pathways connecting distinct regions in phase space.

Intuitively, the role of chaotic dynamics can be viewed as generating "shortcut" trajectories in the semiclassical regime, connecting disparate and typically isolated states. These shortcuts, enabled by coherent quantum interference, allow for an effective entropy reduction as the system preferentially explores fewer, correlated microstates rather than uniformly exploring phase space.

Thus, quantum chaos significantly alters thermodynamic behaviour, enabling selective entropy reduction by leveraging coherence and dynamical tunnelling. This interplay of coherence, interference, and chaotic dynamics provides powerful mechanisms for engineering quantum states with enhanced control over thermodynamic properties.

\section{Quantum and Semiclassical Mechanisms in Biological Systems and Molecular Machines}

Biological systems and molecular machines operate inherently at scales where quantum mechanical effects can significantly influence their thermodynamic behaviours. In particular, coherence and semiclassical chaotic dynamics can guide these systems into specific functional states characterised by reduced entropy. Such states can subsequently be exploited to perform mechanical or chemical work. This chapter provides a detailed mathematical description of how these quantum and semiclassical mechanisms arise naturally in biological contexts, along with specific molecular structures and dynamical conditions that enhance coherence and counter-ergodic behaviour.

Biological molecular machines, such as enzyme complexes, photosynthetic reaction centers, and motor proteins, often operate near the quantum-to-classical boundary. At this boundary, quantum coherence and semiclassical chaotic behaviour become pronounced. Quantum coherence allows for constructive interference among specific transition pathways, enhancing the probability of transitions into thermodynamically improbable, lower entropy states. Semiclassically, chaotic dynamics introduces complex, highly sensitive pathways that amplify quantum coherence and enable chaos-assisted tunneling, effectively linking distant, energetically separated molecular states.

Mathematically, the coherent evolution of quantum states within a molecular machine can be described by the density matrix formalism. The time evolution of the density matrix  under coherent conditions is governed by the von Neumann equation:
\begin{equation}
i\hbar\frac{d\rho}{dt}=[H,\rho]
\end{equation}
where $H$ is the system Hamiltonian. In biological environments, coherence is maintained transiently due to coupling to the surrounding environment. Thus, the total Hamiltonian is typically partitioned into a system part $H_{S}$, an environment part $H_{E}$, and an interaction part 
\begin{equation}
H=H_{S}+H_{E}+H_{SE}
\end{equation}

To describe the environmental effects explicitly and realistically, one often employs the open quantum system framework, leading to the Lindblad master equation:

\begin{equation}
\frac{d\rho}{dt}=-\frac{i}{\hbar}[H_{S},\rho]+\sum_{k}\gamma_{k}(L_{k}\rho L_{k}^{\dagger}-\frac{1}{2}\{L_{k}^{\dagger}L_{k},\rho\})
\end{equation}

where the operators $L_{k}$ represent interaction channels with the environment, and $\gamma_{k}$ are related to the rates of decoherence and dissipation.

To achieve coherence-driven counter-ergodicity, the biological system exploits structured environmental interactions. Natural biological structures such as photosynthetic complexes and protein scaffolds inherently possess vibrational and electronic energy levels tuned to maintain coherence over biologically relevant timescales. Such coherence allows the quantum system to evolve preferentially toward specific states. Formally, this mechanism can be described by the projection onto coherent subspaces
\begin{equation}
\rho_{c}=P\rho P
\end{equation}

where $P$ is a projection operator onto the coherent subspace defined by biologically relevant molecular states.

Chaotic dynamics further enhance coherence through chaos-assisted tunnelling, described by the semiclassical propagator in chaotic environments:

\begin{equation}
K(q_{f},q_{i};t)=\sum_{\alpha}A_{\alpha}e^{i S_{\alpha}(q_{f},q_{i};t)/\hbar}
\end{equation}

where trajectories $\alpha$ explore the chaotic regions of phase space. Chaotic dynamics generate rich interference structures, increasing the amplitude of otherwise improbable transitions by constructive interference, represented explicitly by
\begin{equation}
P_{chaos-assisted}=\Bigg|\sum_{\alpha,\beta}A_{\alpha}A_{\beta}^{*}e^{i(S_{\alpha}-S_{\beta})/\hbar}\Bigg|^{2}
\end{equation}

Specific molecular mechanisms amplify these coherence effects. Examples include resonant energy transfer, vibrational mode coupling, and conformational fluctuations tuned by molecular architecture. For instance, the coupling between electronic excitation states and vibrational modes in photosynthetic complexes can amplify coherence by resonant energy transfer processes described by a Hamiltonian of the form
\begin{equation}
H_{resonant}=\sum_{m,n} J_{m,n}(\ket{m}\bra{n}+\ket{n}\bra{m})
\end{equation}

where $J_{mn}$ are coupling constants and $\ket{m}$, $\ket{n}$ are molecular energy states. Such coupling naturally establishes pathways with enhanced coherence and reduced entropy.

Moreover, coherent environmental noise (structured phonon modes) can further stabilise quantum coherence and prolong its lifetime, described explicitly by coupling Hamiltonians of the form
\begin{equation}
H_{SE}=\sum_{j,k}g_{jk}(\ket{j}\bra{j})\otimes (b_{k}+b_{k}^{\dagger})
\end{equation}

where $b_{k}$, $b_{k}^{\dagger}$ are phonon annihilation and creation operators, and $g_{jk}$ describe the coupling strength. Such structured noise environments effectively guide quantum states toward low-entropy, functionally relevant configurations through counter-ergodic dynamics.

Thus, biological systems inherently leverage quantum coherence and semiclassical chaotic dynamics to facilitate counter-ergodic transitions into lower entropy states, enabling molecular machines to perform mechanical and chemical work efficiently. Understanding and harnessing these principles provide deep insights into biological energy conversion and molecular machinery, potentially inspiring novel bio-inspired quantum technologies.

\section{Conclusion and Applications}

Understanding quantum coherence and chaos-driven semiclassical mechanisms has profound implications for molecular robotics, evolutionary biology, and synthetic biology. Molecular robots designed to exploit coherence and chaotic dynamics can efficiently perform mechanical and chemical work, overcoming traditional thermodynamic limitations. Such coherence-driven molecular machines could revolutionise nano-engineering, enabling precise molecular control and highly efficient energy transformations.

In molecular evolutionary biology, insights into quantum coherence and chaotic semiclassical dynamics offer a deeper understanding of how complex biological systems evolve to optimise efficiency and functionality. Recognising how biological evolution naturally harnesses quantum coherence and chaos-assisted tunnelling could shed light on evolutionary strategies underlying bioenergetics, enzyme catalysis, and motor protein function.

In synthetic biology, incorporating these quantum and semiclassical principles could inspire novel bio-inspired designs, enhancing our ability to engineer biological systems with finely tuned thermodynamic properties. Thus, elucidating quantum coherence and chaotic dynamics paves the way for revolutionary advancements in both fundamental biological understanding and innovative technological applications.

\end{document}